\documentstyle[12pt,epsfig]{article}
\makeatletter

\@addtoreset{equation}{section} \makeatother

\newcommand{\del}{\nabla}
\newcommand{\veca}{{\mathbf{a}}}
\newcommand{\vecb}{{\mathbf{b}}}
\newcommand{\vecx}{{\mathbf{x}}}
\newcommand{\vecy}{{\mathbf{y}}}
\newcommand{\paa}{\partial}
\newcommand{\sqrtg}{\sqrt{-\gamma}}
\newcommand{\si}{{\sigma}}
\newcommand\spr[1]{\mathaccent19{#1}}
\renewcommand\={\!\!\!&=&\!\!\!}

\begin{document}

\renewcommand{\thefootnote}{\fnsymbol{footnote}}

\begin{flushright}
DAMTP-2000-46 \\ IMPERIAL/TP/99-00/27 
\end{flushright}
\vskip 12pt

\begin{center}

{\large\bf 
 Dynamics and properties of chiral
cosmic strings in \vskip 0.1cm 
 Minkowski space}

\vskip 1.2cm
{\large A.C. Davis$^{a}$\footnote{E-mail: {\tt
A.C.Davis@damtp.cam.ac.uk}}, T.W.B. Kibble$^{b}$\footnote{E-mail:
{\tt T.Kibble@ic.ac.uk}}, M. Pickles$^a$\footnote{E-mail: {\tt
M.Pickles@damtp.cam.ac.uk}} and D.A. Steer$^{a}$\footnote{E-mail:
{\tt D.A.Steer@damtp.cam.ac.uk}}}\\ \vskip 5pt \vskip 3pt {\it
a}) D.A.M.T.P., C.M.S., Wilberforce Road, Cambridge, CB3 OWA, U.K.
\vskip 3pt {\it b}) Blackett Laboratory, Imperial College, London, SW7
2BW, U.K.\\
\vskip 0.3cm
\end{center}

\vskip 1.2cm

\renewcommand{\thefootnote}{\arabic{footnote}}
\setcounter{footnote}{0} \typeout{--- Main Text Start ---}

\begin{abstract}

Chiral cosmic strings are produced naturally at the end of
inflation in supersymmetric models where the symmetry is broken via a
D-term.  Consequently in such theories, where both inflation and
cosmic strings contribute to the density and CMBR (microwave
background) perturbations, it is necessary to understand the evolution
of chiral cosmic string networks.  We study the dynamics of chiral
cosmic strings in Minkowski space and comment on a number of
differences with those of Nambu-Goto strings.  To do this we follow
the work of Carter and Peter who showed that the equations of motion
for chiral cosmic strings reduce to a wave equation and two
constraints, only one of which is different from the familiar
Nambu-Goto constraints. We study chiral string loop solutions
consisting of many harmonics and determine their self-intersection
probabilities, and comment on the possible cosmological
significance of these results.

\end{abstract}

\vskip 0.7cm

\section{Introduction}

In the last few years many high accuracy
calculations have been made of cosmological consequences of
Nambu-Goto (NG) cosmic strings \cite{James1,Avelino,CHM,Levon}.
Indeed, such predictions were recently
compared to the Boomerang data \cite{boomerang,Bouchet,Contaldi}.
There are good reasons why most studies of the cosmological effects of
topological defects (see \cite{HK,ViSh} for a summary) have
concentrated on NG strings:  these are the simplest type
of cosmic string, and their equations of motion (at least in Minkowski
space) can be solved exactly.  On a lattice one can use the highly
efficient Smith-Vilenkin algorithm \cite{SV}; and in fact some of the recent
predictions are based on Minkowski space codes of NG network evolution
\cite{CHM,Contaldi}.

However NG cosmic strings, of which the simplest example are the
strings formed in the Abelian Higgs model, are not likely to be
the most realistic type of cosmic string.  Cosmic strings that can
create significant density perturbations require GUT scale physics.
If the Higgs field that forms the string couples to fermions in the
GUT theory --- as might well be expected of a Higgs field --- then
these fermions yield zero modes in the core of the string \cite{witten},
thereby generating a current (which need not be
electromagnetically coupled) along the string.

In this paper we are interested in chiral cosmic strings which
arise naturally in supersymmetric (SUSY) theories \cite{DDT1}, where a $U(1)$
symmetry is broken with a Fayet-Iliopoulos D-term,
resulting in a
single fermion zero mode which travels in only one direction
along the string --- this defines a chiral string.
In the cosmological context, chiral strings are
automatically formed at the end of inflation in SUSY models with a
D-term \cite{rachel}.  Furthermore the zero mode (or chiral nature of
the string) survives the subsequent supersymmetry breaking phase
transition \cite{DDT2}, and consequently both inflation and chiral
cosmic strings contribute to the density and CMBR perturbations in
this scenario.  Calculations of these observable predictions were
carried out recently \cite{Contaldi,CHM2,battye}, and there the $C_l$'s were
decomposed as
$$
C_l = \alpha C_l^{{\rm inflation}} + (1-\alpha)C_l^{\rm NG
\; strings}
$$
with $0 \leq \alpha \leq 1$.
However, we believe that the use of the $C_l$ from NG strings
in the above formula is an oversimplification
at least in the inflation plus chiral cosmic string
scenario mentioned above. In the case of chiral cosmic strings, the
presence of the fermion zero mode is likely to have a significant
effect on the dynamics of the string network, which could therefore
evolve very differently to a NG network.  Indeed the action
describing the evolution of chiral cosmic strings is very different
from the NG action \cite{CP}.  The purpose of this paper is to quantify
some of the differences between the evolution of these two different
types of cosmic string network.

Our starting point is the action for chiral cosmic strings first
proposed in \cite{CP}.  As usual, in order to derive the equations of
motion from this action, gauge choices must be made: in Minkowski
space with suitable gauge choices the equations of motion
reduce to the remarkably familiar form given by \cite{CP}
\begin{eqnarray}
\frac{\paa^2 \vecx}{\paa t^2} - \frac{\paa^2 \vecx}{\paa
\sigma^2} \= 0 \qquad \Longrightarrow \qquad
\vecx(t,\si) = \frac{1}{2}[\veca(t+\si) + \vecb(t-\si)],
\nonumber
\\
\spr{\veca}^2 \= 1,
\nonumber
\\
\spr{\vecb}^2  \!\!\!& \le &\!\!\! 1,
\nonumber
\end{eqnarray}
where for instance $\spr\veca(q)\equiv d\veca(q)/dq.$
These can be recognized as the usual NG equations of motion
\cite{HK} with the only difference being the constraint on the
derivative of $\vecb$ which must now lie {\em within} the
Kibble-Turok sphere rather than on it.   We show in section
\ref{sec:gauge} that the physical reason for this stems from the
conserved charge on chiral strings: if this charge is zero then
$\spr\vecb^2  = 1$ and one is left with NG strings as required.

If the charge on the strings is maximal then $\spr\vecb^2  =0$.  As
we show in section \ref{sec:res}, this latter special case is very
interesting since it implies that $\dot{\vecx} = \vecx'$ with
$|\dot{\vecx}|=1/2$ so that the strings move along themselves at half
the speed of light and never change shape or self-intersect.  In the
case of loops, this solution corresponds to arbitrary-shape stable
vortons.  For infinite strings it means that self-intersections
(which produce loops) never occur.  Since this is an important
mechanism of energy loss in the case of NG strings, this result
already gives an indication that the evolution of chiral and NG
cosmic string networks may be very different.  For general chiral
string solutions, self-intersection is certainly possible, but we may
expect that the probability is lower than in the NG case.  To
quantify the differences between NG and chiral string evolution, we
study in section \ref{sec:res} the self-intersection probability of
loops with different numbers of harmonics on them and different
conserved charges.

The plan of this paper is the following.  For pedagogical reasons, we
begin in section \ref{sec:gen} by briefly reviewing the work of Carter
and Peter (CP) \cite{CP}.  For chiral strings it is not possible to
impose exactly the same gauge conditions as are usually chosen for NG
strings.  In section \ref{sec:gauge} we review the possible gauge
choices, in particular those made by Martins and Shellard \cite{MS} and
by CP.  We follow the latter, whose choice leads to very simple
equations of motion, which are almost identical to those of the NG
string and which, most importantly, are exactly integrable in Minkowski
space.  In section \ref{sec:res} we use these simple equations to
discuss general properties of chiral strings as well as the
self-intersection properties of loops with different numbers of
harmonics and different charges.  Finally, our conclusions and plans
for future work are discussed in section \ref{sec:concl}.

{\em Note:}  Whilst we were trying to extend the results presented
here to FRW universes, a paper by Blanco-Pillado {\it et al.} \cite{BP}
appeared which also obtains the equations of motion above, though from a
rather different point of view.  Here we follow more closely the work
of CP \cite{CP}, and extend both of these papers to study
some general properties of chiral cosmic strings and the
self-intersection of loops.

\section{Chiral and Nambu-Goto strings}

\subsection{Action and equations of motion}
\label{sec:gen}

For pedagogical reasons, we here review briefly the work of Carter and
Peter \cite{CP}.

The action for chiral cosmic strings  they proposed involves a
dimensionless scalar field $\phi$ which can be
interpreted as the phase of the current carriers
condensed on the string.  The action is \cite{CP}
\begin{equation}
S = - \int d^2 \sigma\, \sqrt{-\gamma} \left(m^2 - \frac{1}{2} \psi^2
\gamma^{ij} \phi_{,i} \phi_{,j}\right),
\label{Baction}
\end{equation}
where $\gamma_{ij}$ ($i,j =\{0,1 \}$)  is
the induced metric on the world sheet:
$$
\gamma_{ij} = x_{,i}\cdot x_{,j} \equiv g_{\mu \nu}x^{\mu}_{,i}
x^{\nu}_{,j}.
$$
Here $g_{\mu \nu}$ is the background metric, with signature
$(+,-,-,-)$ (which in the following we shall take to be the Minkowski
metric) and $x^{\mu}(\sigma^0,\sigma^1)$ denotes the position of the
string at world-sheet coordinates $\sigma^{i}$. The first term in
(\ref{Baction}) is just the NG action for a string with tension $m^2$.

The action (\ref{Baction}) is invariant under reparametrizations,
$\si^i\to\tilde\si^i=\tilde\si^i(\si^j)$, and also under
transformations of $\phi$, with a compensating transformation of
$\psi$:
\begin{equation}
\phi\to\tilde\phi(\phi),\qquad{\rm with}\qquad
\psi\to\tilde\psi=\left({d\tilde\phi\over d\phi}\right)^{-1}\psi.
\label{phitrans}
\end{equation}
These freedoms must be removed by making gauge choices, as discussed
below.

The dimensionless Lagrange multiplier $\psi$ sets the constraint
\begin{equation}
\gamma^{ij} \phi_{,i} \phi_{,j} = 0
\label{cpsi}
\end{equation}
which ensures that the cosmic strings are indeed chiral.  Generally,
current-carrying strings are characterized by two currents
and their corresponding charges \cite{ViSh}.
One of these currents, proportional
to $\gamma^{ij}\phi_{,j}$, is conserved by virtue of the equations of
motion; the other, proportional to
$\epsilon^{ij}\phi_{,j}$, is topologically conserved.  However, for
chiral strings, because of (\ref{cpsi}) the two currents coincide, and
so therefore do the two corresponding charges $Z$ and $N$.

In Minkowski space, the equations of motion following from
(\ref{Baction}), in addition to (\ref{cpsi}), are
\begin{equation}
\paa_i (\sqrtg \psi^2 \gamma^{ij} \phi_{,j})=0
\label{currcon}
\end{equation}
and finally
\begin{equation}
\paa_i \left[ \sqrtg \left( \gamma^{ij} + \frac{\psi^2}{m^2} \phi^{,i}
\phi^{,j} \right) x^{\mu}_{,j} \right]=0.
\label{Mink}
\end{equation}

In 1+1 dimensions, a scalar field whose gradient is everywhere null
is necessarily harmonic, i.e., (\ref{cpsi}) implies
$$
0 = \del^j\del_j\phi = \frac{1}{\sqrtg}\paa_i(\sqrtg
\gamma^{ij}\phi_{,j}),
$$
which is consistent with (\ref{cpsi}) only if $\psi$ is a
function of $\phi$, $\psi=\psi(\phi)$.  We shall verify this later in
particular coordinate systems.

From (\ref{currcon}), we may provisionally define the current as
$j^i=\psi^2\phi^{,i}$, which is conserved and null, satisfying
$$
j_i j^i = 0.
$$
However, this expression is \emph{not} invariant under the $\phi$
transformation (\ref{phitrans}).  There is an ambiguity in the
definition of $j^i$: {\it any} current of the form
$j^i=f(\phi)\phi^{,i}$ is null and conserved; there is an infinity of
conservation laws.  This is a peculiarity of null currents in 1+1
dimensions.  For definiteness, we choose the invariant current
\begin{equation}
 j^{i} = \psi \phi^{,i}.
\label{currT}
\end{equation}

\subsection{Gauge choices}
\label{sec:gauge}

To proceed further, gauge choices must be made.  The action
(\ref{Baction}) is reparametrization invariant --- we can replace
$\si^0$ and $\si^1$ by any functions of these variables.  Moreover,
we can replace $\phi$ by any function of itself, provided we change
$\psi$ to compensate as in (\ref{phitrans}).

For the NG string it is usual to choose the conformal gauge in which
$\gamma_{ij}(\si^k) = \Omega(\si^k)\eta_{ij}$, with $\eta_{ij}={\rm
diag}(1,-1)$.  Explicitly, if we write $\tau=\si^0$, $\si=\si^1$, and
denote derivatives with respect to $\tau$ and $\si$ by a dot and
prime respectively, the conformal gauge is specified by two
conditions:
\begin{equation}
\dot x^2 + x'^2 = 0
\label{confone}
\end{equation}
and
\begin{equation}
\dot x\cdot x' = 0.
\label{conftwo}
\end{equation}
This implies that the string's velocity is perpendicular to its
tangent vector.  For the NG string, because of the conformal
invariance of the action,
these two conditions are not in fact independent, and therefore
do not fully specify the coordinates.  We can in addition impose the
temporal gauge condition
\begin{equation}
\tau = t \equiv x^0.
\label{temporal}
\end{equation}
For chiral strings, however, these three conditions are inconsistent,
so different choices are needed.

We first recall the choices made by Martins and Shellard (MS) \cite{MS},
who studied a number of properties of chiral cosmic string loops.
They opted to maintain the temporal gauge condition
(\ref{temporal}).  As noted above, there is no longer freedom to choose
the full conformal gauge as well.  Instead MS chose the
world-sheet metric to be diagonal, maintaining (\ref{conftwo}) but not
(\ref{confone}).
On defining $\epsilon^2 =
(\sqrtg \gamma^{00})^2 = {\vecx'^2}/(1-\dot{\vecx}^2)$, it
follows from (\ref{cpsi}) that $ \phi'^2 = \epsilon^2 \dot{\phi}^2
$ so that equation (\ref{currcon}) yields
$$
\paa_t[\psi^2 \phi'] = \paa_{\si} [\psi^2 \dot{\phi}],
$$
confirming the general result that $\psi = \psi(\phi)$.  MS chose to
fix $\phi$ by setting  $\psi^2 =$ const $=1$. The equations of motion
following from (\ref{Mink}) are then given by \cite{MS}
\begin{eqnarray}
[\epsilon (1 + \Phi)]\dot{\phantom\imath} \= \Phi'
\nonumber
\\
\epsilon(1 + \Phi)\ddot{\vecx} \= \left[
(1-\Phi)\frac{\vecx'}{\epsilon} \right]'+ \dot{\Phi}\vecx'+ 2 \Phi
\dot{\vecx}'
\label{MSeqs}
\end{eqnarray}
where $\cdot=d/dt$ and $'=d/d\sigma$ and 
$\Phi = \dot{\phi}^2/(m^2 \gamma_{00})$.
Loop solutions to these equations were studied in \cite{MS,MS2}.

Since these gauge choices lead to rather complicated equations of
motion, we shall opt instead to follow the paper of Carter and
Peter (CP) \cite{CP} and choose one of the world-sheet coordinates to
be proportional to $\phi$.  That is, choose $\eta = m^{-1} \phi$ (the
factor of $m^{-1}$ is introduced for dimensional reasons) to be one
world-sheet coordinate and denote the second by $q$.  By
(\ref{cpsi}) this implies
\begin{equation}
\gamma^{\eta \eta} = 0 \qquad \Longrightarrow  \qquad
\gamma_{qq}=0.
\label{c1}
\end{equation}
Thus the line element on the world-sheet is
$$
ds^2 = A d\eta^2 + 2 \Omega dq d\eta
$$
where
$$
\Omega \equiv \gamma_{\eta q} = \sqrtg = x_\eta\cdot x_q
$$
and
\begin{equation}
A \equiv \gamma_{\eta \eta} = x_{,\eta} \cdot x_{,\eta}.
\label{Adef}
\end{equation}
With this choice of coordinates, we see again that $\psi = \psi(\phi)$
since equation (\ref{currcon}) gives
$$
0 = \paa_q[\sqrtg\gamma^{q\eta}m^{-1} \psi^2] =
m^{-1}\paa_q [\psi^2]
\qquad\Longrightarrow\qquad \psi =  \psi(\phi).
$$
We also note from (\ref{currT}) that
\begin{equation}
j_\eta=m\psi,\quad j_q=0,\qquad\Longrightarrow\qquad
j^\eta=0,\quad j^q=\frac{m\psi}{\Omega}.
\label{chicurr}
\end{equation}

Now, the equation of motion (\ref{Mink}) gives
\begin{equation}
2 \paa_q \paa_{\eta} x^{\mu} + \paa_q [F(\paa_q x^{\mu})] = 0
\label{mot1}
\end{equation}
where
$$
F =
m^2(\Omega \gamma^{qq} + \psi^2 \gamma^{q\eta})
= \frac{m^2}{\Omega}
\left( \psi^2 - A \right).
$$
It is now clear that we can further simplify the equations
(\ref{mot1}) by choosing the second coordinate $q$ in such a way that
$F=0$.  Happily this is a consistent choice because then
$A = x_{,\eta} \cdot x_{,\eta} = \psi^2$ should be a function of
$\phi$ only, independent of $q$.  But this is indeed the
content of the simplified equation of motion, (\ref{mot1}) with $F=0$,
namely
\begin{equation}
\paa_q \paa_{\eta} x^{\mu} = 0.
\label{finalmink}
\end{equation}
The simplicity of this equation shows the convenience of this gauge
choice, in which both $\eta$ and $q$ are characteristic
coordinates.  The general solution of the equation of motion is
$$
x^{\mu}(q,\eta) = \frac{1}{2}[a^{\mu}(q) + b^{\mu}(\eta)],
$$
exactly as for the NG string.

Within the gauge choices so far made, we still have freedom to
transform each of the coordinates $\eta$ and $q$ separately: $\eta
\to\tilde{\eta}(\eta)$ and $q \to\tilde{q}(q)$.  It is convenient to
choose them so that $\eta = a^0$ and $q = b^0$, and hence $t\equiv x^0
= {1\over2}(\eta+q)$.  This is essentially a temporal gauge.
Finally, let
$$
q=t+\si, \qquad \eta = t-\si.
$$
Then the equations of motion (\ref{finalmink}) and
constraints (\ref{Adef}) and (\ref{c1}) reduce to
\begin{eqnarray}
\ddot\vecx - \vecx'' \= 0 \qquad \Longrightarrow \qquad
\vecx(t,\si) = \frac{1}{2}[\veca(q) + \vecb(\eta)],
\nonumber
\\
\left( \frac{d\veca}{dq}\right)^2 \!\!\!&\equiv&\!\!\!
\spr{\veca}^2 =1,
\label{eqnstosolve}
\\
\left(\frac{d\vecb}{d\eta}
\right)^2 \!\!\!&\equiv&\!\!\! \spr{\vecb}^2  \le 1,
\nonumber
\end{eqnarray}
where $\cdot = d/dt$ and $'=d/d\sigma$.  
Notice that equations (\ref{eqnstosolve}) resemble very closely the
NG equations and constraints in the temporal, conformal gauge:  the
only difference is that now $\spr{\vecb}$ is constrained to lie {\em
within} the Kibble-Turok sphere rather than on it.  The physical
reason for this will be discussed below.

We note that Blanco-Pillado {\it et al.} \cite{BP} made essentially the
same gauge choice, though without introducing the Lagrange-multiplier
variable $\psi$.

The meaning of the coordinate $\sigma$ can be understood by
constructing the stress energy tensor.  With the gauge choices made
above this is given by
$$
T^{\mu \nu} (t,\vecy)= m^2 \int d\sigma \left( \dot x^{\mu} \dot
x^{\nu}  - x^{\mu}{}'x^{\nu}{}' \right) \delta^3(\vecy -
\vecx(t,\sigma)),
$$
which
 is formally identical to the NG stress energy tensor in
the conformal-temporal gauge.  Since
$$
T^{00} (t,\vecy)= m^2 \int d\sigma\, \delta^3(\vecy -
\vecx(t,\sigma)) 
$$
is conserved in Minkowski space, it follows that $\sigma$ again
measures the energy or `invariant length' along the
string.

Finally, we examine the reason for the inequality $\spr{\vecb} ^2 <
1$, which follows from the conserved charge on the string.  From
(\ref{chicurr}) the physical current on the string is given by
$$
 j^t = j^{\sigma} =\frac{m \psi}{2\Omega},
$$
so that the conserved charge is
$$
N = Z = \int d \sigma\, \sqrtg j^t = \frac{1}{2} \int d \sigma\, m\psi.
$$
Now, let
$$
\spr{\vecb} ^2 = k^2,
$$
so that $\psi^2= A = x_{,\eta} \cdot x_{,\eta} = \spr b^2/4 = (1 -
k^2)/4$, from which it clearly follows that $k^2 \le 1$. Note that
\begin{equation}
N = \frac{m}{4} \int d \sigma\,\sqrt{1-k^2},
\label{Ndef}
\end{equation}
so the value of $k$ determines the charge on the string:  this
takes its maximum value when $k=0$ everywhere (as we will see below
this corresponds to interesting vorton solutions), and $N=0$ when
$k\equiv1$, which is exactly the NG limit as required.

In the next section we study the self-intersection properties of
loops with different values of $N$.

\section{Properties of chiral cosmic strings and loop
self-intersections}\label{sec:res}

In this section we first describe some general properties of chiral
cosmic strings which follow from equations (\ref{eqnstosolve}).
Loop self-intersections are then studied.

\subsection{General properties}
From equations (\ref{eqnstosolve}), the velocity and tangent vectors of
the string are given by
\begin{equation}
\dot{\vecx}(t,\si) = \frac{1}{2}[ \spr{\veca} + \spr{\vecb} ],
\hspace{1.2cm}
{\vecx'}(t,\si) = \frac{1}{2}[ \spr{\veca} - \spr{\vecb} ].
\label{vectang}
\end{equation}
Here $|\spr\veca|=1$, while $|\spr{\vecb}| = k \le 1$.  We shall
generally assume that there is a nonzero current, so that $k<1$.  It
then follows that chiral current-carrying cosmic strings in Minkowski
space cannot have zero velocity.  (Thus stationary loops do not exist
for example.) Similarly the tangent vector of the string never
vanishes either so that there are no cusps on these strings.

From equations (\ref{vectang}) it also follows that
$$
\dot{\vecx}\cdot{\vecx'} = \frac{1}{4}[1-k^2] > 0 .
$$
so that the velocity of a point on the string is not perpendicular to
its tangent vector.  Notice that this result is due to the gauge
conditions we have chosen:  with the gauge choice of MS,
$\dot{\vecx}\cdot{\vecx'} = 0$.

Observe also that when $k=0=|\spr{\vecb}|$, equation (\ref{vectang})
implies that $\dot{\vecx} = {\vecx'}$.  Since their velocity is
always parallel to the tangent vector, these strings
do not change their shapes, and thus never self-intersect.
Furthermore, given that when $k=0$, $\dot{\vecx}^2 + {\vecx}^{'2} =
\frac{1}{2}$ and $\dot{\vecx}\cdot{\vecx'}=\frac{1}{4}$, it follows
that for
$k=0$
$$
|\dot{\vecx}|=\frac{1}{2} = |{\vecx'}|.
$$
Thus the strings move at half the speed of light.  Recall from
(\ref{Ndef}) that when $k=0$ the strings carry the maximal charge.

\subsection{Loops}
We now turn to the properties of loops which must satisfy the
periodicity conditions
$$
\vecx(t,\si+L)=\vecx(t,\si).
$$
From (\ref{vectang}), this implies, in the centre-of-mass frame,
$$
\veca(q+L)=\veca(q),\qquad \vecb(\eta+L)=\vecb(\eta).
$$
It follows that, as in the NG case, the motion of chiral cosmic string
loops is periodic, with period $L/2$ (because
$\vecx(t+L/2,\si+L/2)=\vecx(t,\si)$).  As noted above, for
$k=0$ these loops do not self-intersect and hence are vorton solutions
\cite{rick}.  The majority of studies of vortons to date have assumed that
these are circular loops \cite{rick} (see however
\cite{BrandonNato}).  The vortons with $k=0$ found here
have entirely arbitrary shapes.

For arbitrary $k$ it is possible to construct
solutions of (\ref{eqnstosolve}) just as in the case of NG strings
\cite{neil&tom,Neil,CDH,DES}.  For example, a 1-harmonic loop solution with constant
$k$ is given by
$$
\veca(q) = (\cos q , \sin q, 0)\; ; \qquad \vecb(\eta)=(k\cos \eta,
-k \sin \eta, 0) .
$$
This is a circular string oscillating between
maximum and minimum radii of $(1+k)/2$ and $(1-k)/2$.  (Such a
solution was considered numerically in \cite{CPG} for arbitrary current
carrying loops.) This loop
never self-intersects for any value of $k$.  Higher order
harmonic solutions can also be constructed along very similar lines to
references \cite{Neil,CDH,DES}.

Since loops with $k=0$ never self-intersect, it is interesting to ask
how the  self-intersection probability of a loop with a given number
of harmonics depends on $k$ (or equivalently on the conserved charge
$N$ given in (\ref{Ndef})).\footnote{Whilst initially static NG loops
always self-intersect \cite{neil&tom}, chiral cosmic strings can never be
static as observed above.}  For simplicity, we will study this
question for $k(\phi)$ = constant.  In that case, $N$ is given by
\begin{equation}
N = \frac{m}{4} L \sqrt{1-k^2}
\label{Ndef2}
\end{equation}
where $L$ is the invariant length of the loop.  We now show that
self-intersecting loop solutions exist for $k>0$ through the
construction of an explicitly self-intersecting loop.  Then the
probabilities of self-intersection will be studied numerically for
loops of a fixed invariant length $L$ but with different numbers of
harmonics on them and different values of $k$.

The condition that a loop self-intersects at time $T$ is that there
exists a solution of
\begin{equation}
\veca( T+\sigma_1)+\vecb(T-\sigma_1) =
 \veca(T+\sigma_2)+ \vecb(T-\sigma_2)
\label{sisec}
\end{equation}
for some $0< \sigma_1 \neq \sigma_2 < L$.  To show that
self-intersection is possible, consider the following solutions for
$\veca$ and $\vecb$ that satisfy (\ref{eqnstosolve}):
\begin{eqnarray}
\veca(q) \= \frac{1}{m}(\cos m q, \sin m q,0)
\nonumber
\\
\vecb(\eta)\=\frac{k}{n}(\cos n \eta, \cos \chi \sin n\eta, \sin \chi
\sin n\eta) ,
\label{trial}
\end{eqnarray}
where $n$ and $m$ have no common factors and
$\chi$ is an arbitrary angle.
Now let $c=(\si_1 + \si_2)/2$, $\delta = (\si_1 - \si_2)/2$,
$q = T + c$ and $\eta = T-c$.  Then the
self intersection condition (\ref{sisec}) becomes
$$
\veca(q+\delta ) - \veca(q-\delta) =
\vecb(\eta+\delta)- \vecb(\eta-\delta)
$$
for which we must find solutions for $\eta,q,\delta$.
On substitution of (\ref{trial}), this condition becomes
\begin{eqnarray}
\lefteqn{\frac{1}{m}(-\sin m q \sin m \delta,\cos mq \sin m\delta,0)}
\nonumber
\\
\=
\frac{k}{n}( -\sin n\eta \sin n \delta,\cos n\eta \sin n \delta
\cos \chi,\cos n\eta \sin n \delta \sin \chi ).
\nonumber
\end{eqnarray}
Hence the requirement is that
\begin{equation}
\cos mq = \cos n\eta = 0 \qquad \Longleftrightarrow \qquad
 \sin n\eta = \pm1 = \pm \sin mq .
\label{si1}
\end{equation}
where $\delta$ must satisfy
\begin{equation}
\frac{\sin m \delta}{m} = \pm \frac{k}{n}\sin n \delta.
\label{si2}
\end{equation}
Generically, there are solutions to equations (\ref{si1})--(\ref{si2}),
and hence self-intersections.

\subsection{Numerical study of loop self-intersections}

More generally one can search for self-intersections numerically and
try to determine the self-intersection probability as a function of
$k$ and the number of harmonics on the loop.  To
do this, we used a modified version of the code written by
Siemens and Kibble \cite{SK} to search for self-intersections of 
NG loops.  These
authors built on work of Brown and DeLaney \cite{BD,BCD} who devised a
method of generating odd harmonic series satisfying a given
constraint in terms of products of rotations.  The only difference
between the NG and chiral cosmic string loops is that for the former,
the constraint is $|\spr{\vecb}|^2 = 1$ whilst for the latter the
constraint is $|\spr{\vecb}|^2 = k^2 < 1$.  Thus here we carry out a
simple extension of the work of Siemens and Kibble to study the
self-intersection properties of $M/P$ harmonic loops (the notation
means that there are $M$ harmonics in the solution of
$\veca$, and $P$ in the solution for $\vecb$.)

For more technical details on the code, the reader is referred to
\cite{SK}.  In the results presented in figures \ref{string1}-\ref{string3} below,
the rotation angles were given a uniform
distribution, with the
number of points along the string chosen to be $K=600$.  This gives a
resolution of 0.0104712 radians. The cutoff, below which self-intersection
was not tested, was taken as 0.084 radians corresponding to 8 step
lengths. These are the same parameters as those chosen in \cite{SK}
which have already been seen to work well. Furthermore, decreasing $K$ or
increasing the cutoff did not affect our results.

The self-intersection probability was calculated for $M/M$ cosmic string
loops as a function of $k$ (which corresponds to different
charges on the loop through (\ref{Ndef2})).
Figures~\ref{string1}-\ref{string3} plot
the intersection probability against $\sqrt{1-k^2} \propto N$ for
these $M/M$ harmonic
loops.  (Error bars are one
standard deviation.)
Notice that for $k=0$ the loops do not self-intersect as was already
proved above, whereas for a relatively large range of
$k$ the probability is the same as the NG ($k=1$) case.  Thus
charges on chiral cosmic string loops appear only to have a significant
effect on the dynamics of the loops when these charges are large.
Indeed, if the loops are formed with large charges, they will scarcely
ever intersect and this will lead to a cosmological catastrophe since
the loops (vortons) will dominate the energy density of the universe.
It therefore
remains to understand whether or not these charges are expected to be
large or small when the loops form:  we leave a discussion of this
question to the conclusions.

Finally, we note that while the plots show results for $M/M$
harmonic strings, we also ran the code for strings with different
numbers of  left and right-moving harmonics. This did not
substantially change the intersection probability from that of a $M/M$
harmonic string if the harmonics were both close to $M$.

\begin{figure}
\centerline{\epsfig{file=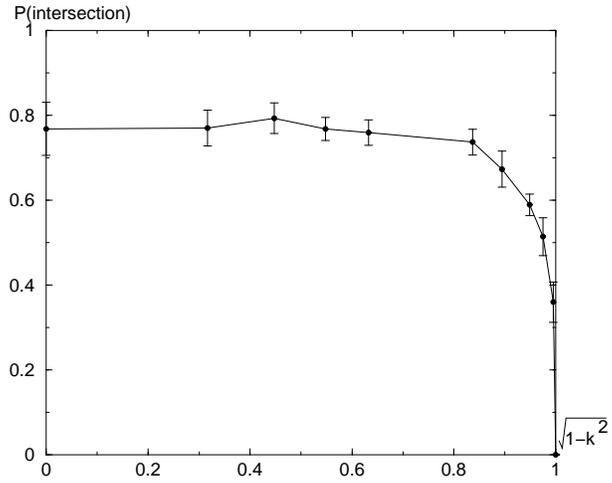,width=8 cm,angle=0}}
\caption{5-5 harmonic string} %
\label{string1}
\end{figure}
\begin{figure} \centerline{\epsfig{file=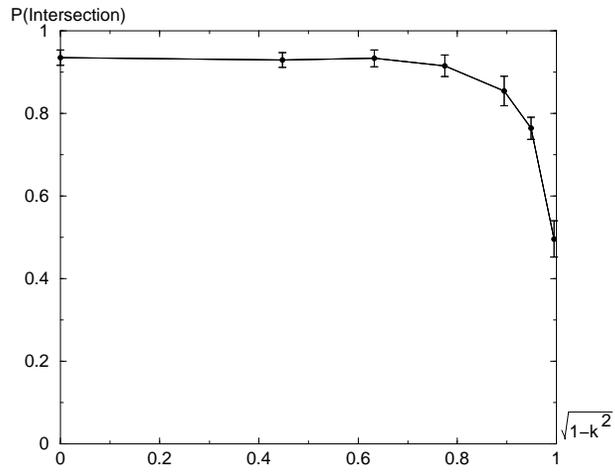,width=8
cm,angle=0}} \caption{11-11 harmonic string} \label{string2}
\end{figure}
\begin{figure}
\centerline{\epsfig{file=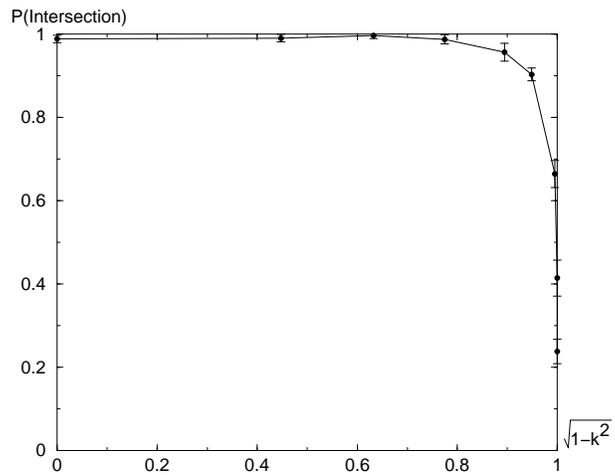,width=8 cm,angle=0}}
\caption{25-25 harmonic string}
\label{string3}
\end{figure}

\section{Conclusions and discussion}
\label{sec:concl}

The basis for this paper was the action (\ref{Baction}) for chiral
cosmic strings first proposed in \cite{CP}.  This is a well defined,
unique, action for strings carrying massless zero-modes which
travel in one direction along the string at the speed of light.
In section \ref{sec:gauge} we reproduced the results of \cite{CP}
showing how, with suitable gauge
choices and treatment of the Lagrange multiplier $\psi$, the
resulting equations of motion are integrable and reduce to the
familiar wave equation with two constraints (\ref{eqnstosolve}).
These two constraints are that $|\spr{\veca}|=1$ and $|\spr{\vecb}|
= k^2 \le 1$.  We noted that the reason why $|\spr{\vecb}|$ lies
within the Kibble-Turok sphere rather than on it is that the
chiral strings carry a conserved charge (associated with the current
on them).  In the limit of zero charge, the equations of motion and
constraints (\ref{eqnstosolve}) reduce to those of NG strings
in the conformal-temporal gauge as required.

We placed a certain emphasis on gauge choices in section
\ref{sec:gauge} since, as we showed in that section, the same
action with less appropriate gauge choices can lead to much more
complicated equations of motion which are not readily integrable, as
for the equations of motion (\ref{MSeqs}) derived in \cite{MS}.

In section \ref{sec:res} we showed that chiral current-carrying cosmic
strings cannot have cusps on them.  Since cosmic rays are
predominantly produced at cusps on NG strings \cite{cosmicrays}, it is likely
that a network of chiral cosmic strings will produce fewer cosmic
rays.

We also showed that when the charge on the string is maximal
(equivalently $k=0$), $\dot{\vecx} = \vecx'$ so that the strings
move along themselves at half the speed of light and never
self-intersect.  In the case of loops these correspond to stationary
vorton solutions of arbitrary shape. For infinite strings it means
that these can never self-intersect to form loops (at least in
Minkowski space).  Since this is the main mechanism for removing
energy from NG string networks, these results suggest that networks
of chiral cosmic strings may evolve very differently from NG cosmic
string networks.

In another step to study the evolution of chiral cosmic string
networks, we considered the self-intersection probability of loops
with $0<k<1$ and different numbers of harmonics (section
\ref{sec:res}).  The results show that only when the charge on the
loop is relatively large does the self-intersection probability
differ significantly from the NG one.

As a result of this work, we are left with a number of important
questions to study in the future.  Maybe the most significant one
of these is to understand what initial value of $k$ might be
expected for the loops and infinite strings formed at the phase
transition (which could be, say, at the end of inflation as
discussed in the introduction).  If $k$ is initially very small
(i.e.\ the charge on the strings is initially close to being
maximal) then chiral cosmic strings are already ruled out, as is
the mixed scenario of D-term inflation and strings \cite{CHM2}.
The reason is that if the chiral cosmic strings effectively never
self-intersect they rapidly come to dominate the energy density of
the universe.
Indeed, since the fermions are traveling in one direction only in
the chiral case, the current and corresponding charge are larger
than in the non-chiral case.
Consequently, we would expect the charge to be
close to maximal and hence $k$ to be small when the
fermion zero modes condense on the string at formation.
If however, the zero modes are formed at a subsequent phase transition,
then $k$ is likely to be closer to unity. Indeed, this is the assumption
made in \cite{CD}, where theories giving rise to chiral cosmic strings
were constrained by the requirement that they should not
over produce vortons.
We have arguments suggesting that $k$ is in fact
initially small; these will be presented elsewhere \cite{DKPS2}.

Here we have restricted attention to loops in which $k$ is constant,
but in fact one should also examine the more general case where $k$
is a function of $\phi$, though always restricted to the range $0\le
k\le 1$.

Another objective would be to try to solve equations
(\ref{eqnstosolve}) in a very similar way to the Smith-Vilenkin
algorithm which is an exact numerical algorithm for solving the
corresponding equations for NG strings in Minkowski space.  However,
the Smith-Vilenkin algorithm is no longer exact for the chiral string
equations: because of the constraint $|\spr{\vecb}|=k<1$, the
vertices will generally not remain on the lattice as the system
evolves.  Indeed we believe that there is no value of $k$ for which
the algorithm can be made to work --- except perhaps $k=0$, a case
that is uninteresting in this context as we already know that there
the strings are effectively stationary.

Finally, one should also consider to what degree the effects of
friction on the evolving chiral cosmic string network are
important. Frictional effects on NG and chiral strings 
are likely to be similar 
since there are no long range forces in either case; this is
unlike the situation for electromagnetically coupled strings \cite{kostas}.
Ultimately the effect of expansion should be
incorporated too, though as in the NG case many predictions can be
made from Minkowski space results \cite{CHM}.  We are currently
studying a number of these questions \cite{DKPS2}.  Our general
conclusion of this paper would be, however, that we have found
evidence to suggest that chiral cosmic string networks evolve very
differently from NG networks.  Hence their cosmological
consequences will be very different, and so some caution
should be used before simply adding the effects of inflation and NG
strings as in \cite{CHM2,battye}, especially when the specific model under
consideration actually produces chiral cosmic strings.

\section*{Acknowledgements}

We thank B. Carter and P. Peter for useful
discussions.  This work is supported in part by PPARC, UK and by an ESF
network.

\typeout{--- No new page for bibliography ---}

\end{document}